\documentclass{ws-ijmpb}
\usepackage{graphicx}
\usepackage{xcolor}
\usepackage[verbose]{hyperref}
\usepackage{physics}
\usepackage{latexsym}
\usepackage{amssymb}
\usepackage{multirow}
\usepackage{amsmath}
\usepackage{float}
\usepackage{amssymb}
\usepackage{hyperref}
\usepackage{subfigure}
\usepackage{subcaption}
\usepackage{caption}
\usepackage{tikz}
\usepackage{lipsum}
\usepackage{latexsym}
\usepackage{amssymb}
\usepackage{amsmath}
\usepackage{amssymb}
\usepackage{tikz}
\usepackage{float}
\usepackage{booktabs}
\hypersetup{colorlinks=false,allbordercolors=blue,pdfborderstyle={/S/U/W 1}}

\begin{document}

\catchline{1}{1}{2026}{}{}
\markboth{Sushita Sharma and Vinod Ashokan}{Computational Modeling of Sublattice Magnetization}

\title{Computational Modeling of Sublattice Magnetization in Co–Tb/Co/Co–Tb Magnetic Multilayer Nanojunctions Between Cobalt Leads}

\author{Sushita Sharma}
\author{Vinod Ashokan\footnote{Corresponding author}}

\address{Department of Physics,\\ Dr. B. R. Ambedkar National Institute of Technology, \\Jalandhar (Punjab) 144 008, India\\
	\email{ashokanv@nitj.ac.in}}

\maketitle


\begin{abstract}
	In this work, we study the sublattice nano-magnetic behavior of $[{\rm Co}_{1-c}{\rm Tb}_c]_{l'}[{\rm Co}]_{l}[{\rm Co}_{1-c}{\rm Tb}_c]_{l'}$ where the unprimed subscript $l$ denotes
the number of pure Co (0001) basal planes and the primed subscript $l'$ denotes the number of Co$_{1-c}$Tb$_{c}$ alloy (0001) basal planes layered ferrimagnetic nanostructures sandwiched between semi-infinite cobalt leads. To perform the analysis, we use the Ising spin-based effective field theory (EFT), which enables us to determine the relevant magnetic exchange parameters for elemental cobalt $J_{\rm Co-Co}$ and terbium $J_{\rm Tb-Tb}$. The exchange interaction parameters are determined by fitting the magnetization calculated using the Ising effective-field theory to the experimentally measured temperature-dependent magnetization curves. This fitting procedure yields the exchange constants for Co and Tb, along with their respective Curie temperatures. We use the magnetic exchange parameter obtained using Ising EFT-based method is then seeded on to mean field theory (MFT) to calculate the sublattice magnetizations of \([{\rm Co}_{1-c}{\rm Tb}_c]_{l'}[{\rm Co}]_{l}[{\rm Co}_{1-c}{\rm Tb}_c]_{l'}\) arranged between semi-infinite cobalt leads. The sublattice magnetizations  for every atomic layer and corresponding lattice sites in the layered ferrimagnetic nanostructures are computed separately. This allows us to analyze spin-site-specific parameters, magnetic moments, magnetization, and ferrimagnetic compensation behavior for varying eutectic compositions, particularly for \(c \leq 0.5\). The results give enhanced insight into magnetic phenomena at the local level and their general relevance in the ultra-thin layered ferrimagnetic structure dynamics confined between semi-infinite cobalt electrodes. These results are needed for the understanding of magnon transport phenomena and consequent magnetic amplification effects and are crucial to the development of spintronic technologies. Most importantly, to date, no first-principles or density functional theory research exists for sublattice magnetization in such multilayered systems. The versatile approach established in this work can also be applied to other nanostructured magnetic systems.
	
\end{abstract}


   \section{Introduction}
	In recent years, there has been much interest in multilayered nano-junctions. These systems have unique properties and have technological applications in the areas of spintronics\cite{serga2007,zutic2004}and magnonics \cite{frackowiak2022magnetic,kruglyak2010,lee2008,schneider2008}.
	In the physics of nano-junctions, these multilayered alloys play a very dominant role
	to classify the sublattice magnetization and atomic interfaces.
	Rare earth - transition metal (RE-TM) \cite{andres2000,anilturk2003} with the magnetic effects of the ferrimagnetic alloy nanostructure has been described in literary works in which interactions such as RE-RE, TM-TM, which involve magnetic moments in sublattices, and even RE-TM  inter-sublattice interactions have been discussed \cite{gonzalez2004,hansen1989,sublatticemag}. The exchange of 3d spins is directly influenced by TM-TM interactions. On the other hand, it is often assumed that RE-RE interactions are indirect, maybe occurring via the 4f-5d-5d-4f span. A similar view has been taken on RE-TM interactions, which are also indirect and result from the combination of interatomic 4f-5d and interatomic 5d-3d \cite{due1993exchange} exchange interactions. Previous studies have shown that alloy interfaces can greatly influence their magnetic properties
	\cite{lamellar,ashokan2014,ashokan2015modeling}.
	
	In this paper, we are dealing with the Co-Tb alloy system in which Co is a transition element and Tb is a rare earth element. At room temperature, Co is ferromagnetic, while Tb is paramagnetic, and they possess Curie temperatures of approximately 119.4 meV and 18.9 meV, respectively. 
	The construction of nano-magnetic nano-junctions with the help of a Co-Tb alloy system in varied configurations can have a wide range of applications in various fields of technology. If we talk about bulk Co${_{1-c}}$Tb${_{c}}$ alloy (where $c$ is  concentration), then it has a wide range of usage in the realms of magnetic storage and optical magnetic switching devices \cite{ciuciulkaite2020magnetic}. A significant understanding of lamellar Co/Tb multilayered alloy nano-junctions is acquired. 
	Through a deep understanding of multilayered alloy nano-junctions of Co-Tb it must be emphasized that while forming Co${_{1-c}}$ Tb${_{c}}$ amorphous alloy, there is the presence of asymmetry in spontaneous diffusion of Co into Tb, which occurs in plane and interfaces at various concentrations. To get control in the diffusion, there are experimental methods which made it possible by defining stable eutectic concentrations ( $c\leq 0.5$ ) without changing the ferrimagnetic nature of the multilayered nanostructure systems \cite{camley1993}. Beyond $c\leq 0.5$ , significant structural changes, enhanced interdiffusion, and possible modifications of the magnetic phase make the present model assumptions less reliable. Our computational formulation itself does not impose any numerical limitation on larger concentrations. Rather, the restriction is motivated by the physical validity of the model and the availability of experimentally relevant exchange parameters. It is important to note that during the formation of the Co$_{1-c}$Tb$_c$ amorphous alloy, an asymmetry is observed due to the natural tendency of cobalt atoms to diffuse into terbium. This diffusion manifests both within the plane and across interfaces, depending on the concentration levels.
	
	Due to different exchange couplings at the interfaces of the bulk being observed, interdiffusion becomes a very crucial part for defining properties of the multilayered nano-junctions. This is very helpful for the determination of nano-magnetic properties \cite{lamellar,ashokan2014} of the multilayered alloy systems. With the help of model calculations, it is noted that there is a modification in the nano-magnetic properties of the multilayered systems when the nanoatomic scale is achieved.

	This work contains a discussion of cobalt-terbium ferrimagnetic nano-junction alloy systems. The nano-junction consists of similar hcp (hexagonal close-packed) planes of cobalt leads and between these similar cobalt leads, there is [Co${_\text{1-c}}$Tb${_\text{c}}]_l$ nano-junction system present with arbitrarily disordered alloy having variation in hcp planes ($l$) with regard of the stable condition of $c\leq 0.5$ in this nano-junction alloy systems. To be capable of executing these calculations, it was necessary to determine the sublattice magnetizations along with the magnetic exchange parameters of the nano-junction alloy configuration. 

    To investigate the magnetic exchange parameters of rare-earth–transition-metal alloys, both in bulk and multilayered systems, several studies have employed in the framework of MFT \cite{camley1988phase,mansuripur1986,vaz2008}. In these approaches, the cobalt spin $S_{\rm Co} $ is often treated as an adjustable parameter, and the model predictions are calibrated to be consistent with MFT-based estimates \cite{demirtas2005}. As a result, the calculated temperature-dependent magnetization can deviate significantly from experimental observations \cite{sublatticemag}, without providing a clear theoretical basis for the chosen spin value. In the present work, we address this issue by determining the exchange constants for Co and Tb directly by fitting the magnetization curve obained from EFT theory with the experimentally measured temperature-dependent magnetization,  the spin values referenced to the fundamental spin states of the respective atoms.The magnetic exchange parameter obtained using Ising EFT-based method is then seeded on to the MFT to calculate the sublattice magnetizations of \([{\rm Co}_{1-c}{\rm Tb}_c]_{l'}[{\rm Co}]_{l}[{\rm Co}_{1-c}{\rm Tb}_c]_{l'}\) arranged between semi-infinite cobalt leads.  This provides a physically motivated description of the magnetic properties of the system.
	
	In this study, modeling of sublattice magnetizations and magnetic exchange constants is achieved with the help of EFT and MFT methods. System of [Co${_{1-c}}\text {Tb}{_{c}}]_{l'}$[Co]${_l}$[Co${_{1-c}}\text{Tb}{_{c}}]_{l'}$ alloy layered system which is being sandwiched in between the semi-infinite cobalt leads at required condition of $c\leq 0.5$ \cite{lamellar}. In the past, single layered alloy nanostructure sandwiched between semi-infinite cobalt leads have been studied \cite{lamellar} therefore, as one would expect, this triple-layered alloy nanostructure system is more complicated than before. Despite their complex nature, these triple-layered systems turned out to have more wider range for the principle of spin wave filtering than single-layered systems, which shows that although triple-layered systems are complex but they have more appositeness than single-layered system. This study is hence motivated with the objective of presenting complex computational modeling for ballistic transport and spin wave scattering (incident from cobalt lead) and primary nano-magnetic properties of this triple nanostructure for basic technological interest. The nanojunction structure is assumed to be in a hcp configuration along the (0001) crystallographic planes that constitute the foundation of these lamellar nanostructures made up of alloy layers ...Co[$l'll'$]Co... and these planes are perpendicular to the crystallographic c-axis, which is parallel to the cobalt lead direction.
	
	This alloy layered nanostructure is approximately 1.5 nm of cobalt terbium alloy system composite, which differentiates it from bulk alloy and system, which consist of multiple layers. The infallible data is required for the computation of magnetic exchange constants and sublattice magnetization of the system, so that modeling studies can be developed for 
	...Co[$[{l'}l{l'}]$]Co... so that ballistic spin wave transport can be discussed. This need is our motivation for EFT calculations so that we can regulate a data which has no fitting parameters by taking the basic spin values (at absolute 0 Kelvin) for cobalt and terbium as $S{_\text{Co}}$=1 and $S{_\text{Tb}}$=3, respectively.
	
	The discussion about the construction of paper is the following, Section 2 consists of EFT ising spin method which is necessary for the computation of reliable values of magnetic exchange constants for pure crystalline materials of cobalt and terbium J${_\text{Co-Co}}$ and J${_\text{Tb-Tb}}$, respectively. In order to get $c\leq 0.5$, the obtained values of exchange constants are allotted to nearest neighbor Co-Co and Tb-Tb interactions that occur in ...Co[$[{l'}l{l'}]$]Co... alloy nano-junction systems. Also, Section 3 comprises combinedEFT ad MFT techniques so that sublattice magnetizations can be calculated for cobalt and terbium sites, temperature-dependent in nature. These locations are situated on hcp basal atomic planes of alloy stacking as $[{l'}l{l'}]$ for lamellar nano-junctions at varying $c$ of alloy with a thickness of $[{2'}2{2'}]$ and $[{3'}3{3'}]$. At the end of this section, sublattice magnetization and correspondent compensation temperatures are also indicated.  In Section 4, the overall summary and conclusions are discussed.
	\section{Ising Effective Field Theory Modeling for Pure Co and Tb Systems:}
	
	\begin{figure}[htbp]
    \centering
    \includegraphics[width=0.8\textwidth]{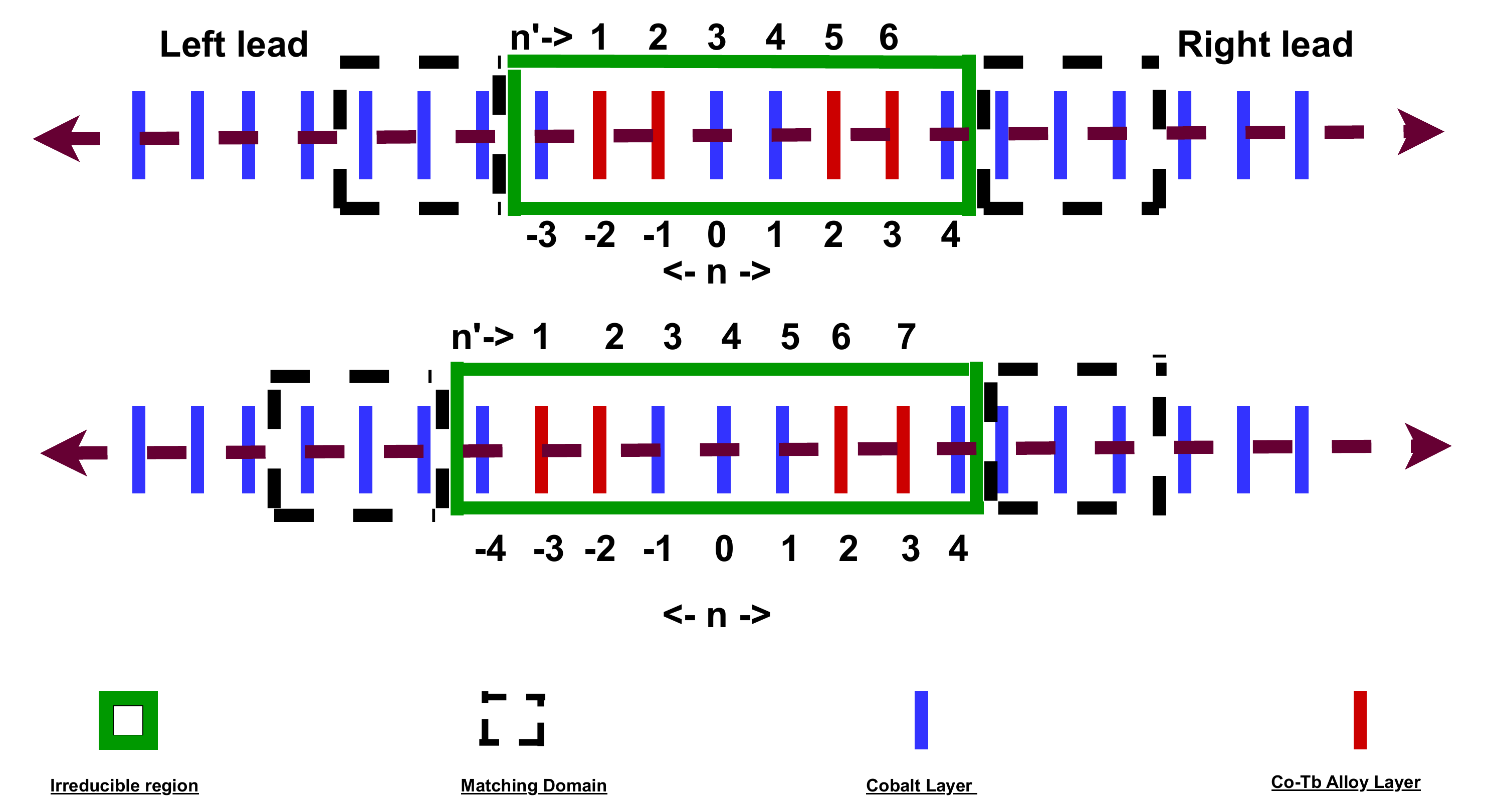} \\
    1(a)\\
    \vspace{2mm}
    \includegraphics[width=0.8\textwidth]{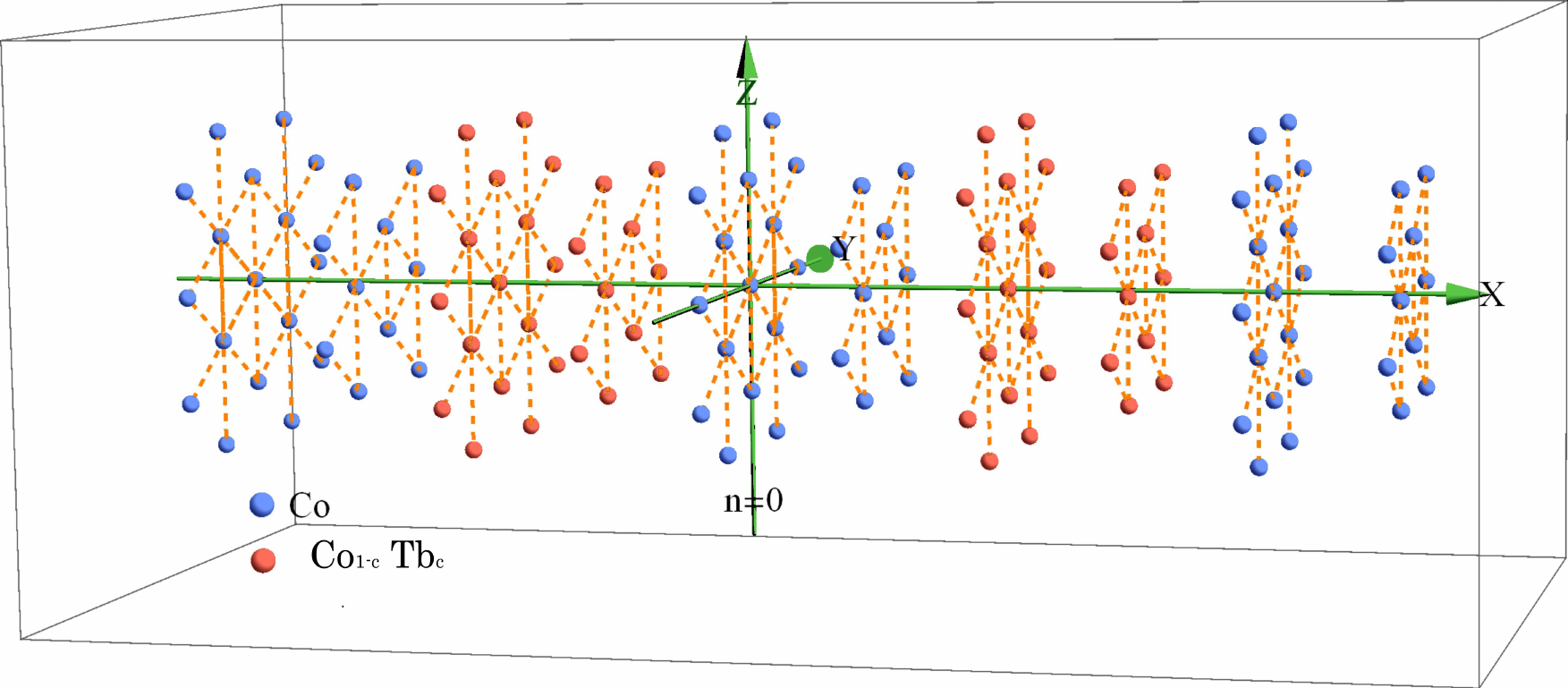}\\
    1(b)
    \caption{Schematic representation of the $[\mathrm{Co}_{1-c}\mathrm{Tb}_c]_2[\mathrm{Co}]_2
[\mathrm{Co}_{1-c}\mathrm{Tb}_c]_2$ nano-junction sandwiched between semi-infinite crystalline cobalt leads. (a) One-dimensional chain representation of the consecutive hcp (0001) atomic planes. (b) Corresponding three-dimensional hcp lattice structure, showing the alternating stacking of cobalt and Co$_{1-c}$Tb$_c$ alloy (0001) basal planes along the $c$-axis, centered at the mid-plane $n=0$. The $c$-axis of the hcp crystal is perpendicular to the (0001) symmetry planes.}
\label{cotb}
\end{figure}
	
	In order to calculate the exchange constants between cobalt-cobalt and terbium-terbium atoms of nano-magnetic system we use Ising effective field theory. This method mainly uses single-site correlation as the key approximation.Decoupling approximation is introduced to simplify the higher-order spin correlation functions appearing in the EFT equations, allowing the coupled nonlinear equations for the layer magnetizations to be solved numerically.\cite{lamellar} The exchange constant calculations for cobalt J${_\text{Co-Co}}$ and terbium J${_\text{Tb-Tb}}$ crystals are extracted by comparing the EFT magnetization output with the experimental data. The calculated results of EFT magnetizations and exchange constants obtained for cobalt are further used for the calculation of sublattice magnetizations for ...Co[$[{l'}l{l'}]$]Co... layered system placed in between the semi-infinite cobalt leads, here the quantity [$l'll'$] with primed $l'$ denote alloy and un-prime denote pure cobalt layers. The schematic representation of multilayered structure of  [Co${_{1-c}}$Tb${_{c}}]{_{2}}$[Co]${_{2}}$[Co${_{1-c}}$Tb${_{c}}]{_{2}}$ which is placed in between semi-infinite cobalt leads are shown in Fig. \ref{cotb}.
	
	\noindent The Ising spin Hamiltonian  ${\mathcal H}$  by neglecting Zeeman effects and local spin anisotropy can  be written as,\cite{gottwald2012asymmetric,strecka2015brief}
	
	\begin{equation}
		{\mathcal H}=-J\sum_{<i,j>} S{_{iz}}S{_{jz}}  .
	\end{equation}
	In the above equation, $J$ is the magnetic exchange constant concerning nearest neighbors that generates order of spin in the selected z direction, sum over nearest neighbors is represented by $<i,j>$. Both Co and Tb have a coordination number $z=12$ with fundamental spins of Co and Tb. The components 
	$S_{iz}$ and $S_{jz}$ refer to the spin 
	S operators at sites $i$ and $j$.
	
	The EFT formulation also involves the estimation of thermodynamic canonical averages, which can be expressed in terms of canonical characteristic functions, or generating functions $f_{Op}$ whose symbolic and numerical expression is given as:
	
	\begin{equation}
		{f}_{Op}(x) = \frac{\text{Tr}(\text{Op.MatrixExp}[-\mathcal{H}_{i}(x)/ k_{\mathrm{B}}T)}{\text{Tr}(\text{MatrixExp}[-\mathcal{H}_{i}(x)/ k_{\mathrm{B}}T])} .
	\end{equation}
	 In above equation, $k$ is Boltzmann Constant, $T$ is temperature and Op is the general spin operator which involves the spin is of the form, \begin{equation}
		\langle O \rangle = \left\langle \left(\exp(JS_z \nabla)\right)^z \right\rangle f_{\text{op}}(x) \big|_{x \to 0} .
	\end{equation} 
    Here $\nabla \equiv \partial/\partial x$ is the differential operator with respect to the auxiliary field $x$.
	The analysis we are conducting focuses solely on the canonical averages of the single-site spin variables because of the decoupling approximation, represented as $\sigma = \langle S_z \rangle $ and $
	q = \langle S_z^2 \rangle.$
	$\mathcal{H}_i(x)$ is now a single site Hamiltonian,
	\begin{equation}
		\mathcal{H}= \sum_{{i}^{}} \sum_{{j}^{}}(-JS_{\text{jz}}) S_{\text{iz}} \equiv \sum_{{i}^{}} \mathcal{H}((x) \equiv 
		\sum_{{i}^{}}-xS_{\text{iz}}.
	\end{equation}
	The spin variables \( \sigma = \langle S_z \rangle \) and \( q = \langle S_z^2 \rangle \) in EFT can be determined using the Van der Waerden identities. For the Co spin \( S = 1 \), VdW identity is:
	\begin{equation}
		\exp(JS_{z} \nabla ) = S_{\text{z}}^{2} \cosh(J \nabla) + S_{\text{z}} \sinh(J \nabla) + 1-S_{\text{z}}^{2} .
	\end{equation}
	The Vdw identity $\exp\left(\frac{J S z \nabla}{2}\right)$ is solved for Tb with fundamental $S=3$ as,
	{\footnotesize
		\begin{align}
			&\exp\left(\frac{J S z \nabla}{2}\right)\nonumber\\ 
			&=  \frac{1}{24}\bigg[280\cosh(u) - 224\cosh(3u) + 6\cosh(5u)\nonumber\\
			&- \cosh(7u)\bigg] + S z \bigg[\frac{1}{720}\bigg(5040\sinh(u) - 3360\sinh(3u) \nonumber\\
			&+ 126\sinh(5u) - \sinh(7u)\bigg)\bigg] + S z^2 \bigg[\frac{1}{36}\bigg(70\cosh(u)\nonumber\\
			&  - 84\cosh(3u) + 15\cosh(5u)- \cosh(7u)\bigg)\bigg]  + S z^3 \bigg[\frac{1}{72}\nonumber\\
			&  \times\bigg(70\sinh(u) - 28\sinh(3u) + 3\sinh(5u) - \sinh(7u)\bigg)\bigg] \nonumber\\
			& + S z^4 \bigg[\frac{1}{72}\bigg(14\cosh(u) + 28\cosh(3u) - 9\cosh(5u)\nonumber\\
			& + 3\cosh(7u)\bigg)\bigg] + S z^5 \bigg[\frac{1}{24}\bigg(28\sinh(u) + 14\sinh(3u) \nonumber\\
			&- 21\sinh(5u) + 7\sinh(7u)\bigg)\bigg] + S z^6 \bigg[\frac{1}{720}\bigg(140\cosh(u) \nonumber\\
			&+ 105\cosh(3u) - 126\cosh(5u) + 35\cosh(7u)\bigg)\bigg] .
		\end{align} 
	}
	In above equation, \begin{equation*}
		u = \frac{J \nabla}{2} .
	\end{equation*}
	
	Our method for calculating spin variables combines symbolic and numerical techniques. In this approach, each variable is associated with numerous contributing terms which grow as the spin value and the number of interacting neighbors (coordination number) increase.
	
	\begin{figure}
		\centering
		\includegraphics[width=0.8\textwidth]{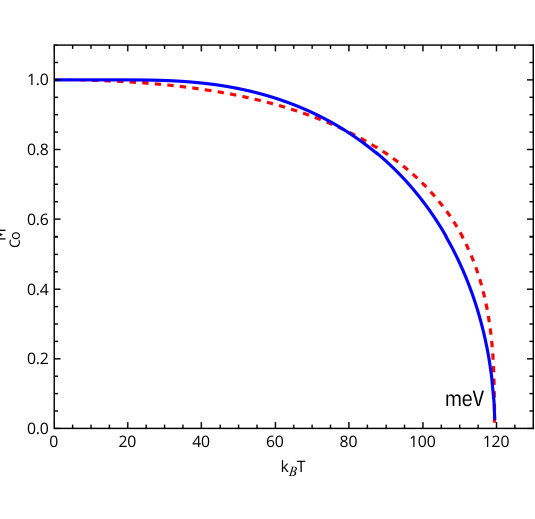}
		\caption{EFT calculated normalized magnetizations (continuous blue curve) for $hcp- Co$ across varying temperatures compared with experimental measurements (dashed red curve)}
		\label{eftco}
	\end{figure}
	
	\begin{figure}
		\centering
		\includegraphics[width=0.8\textwidth]{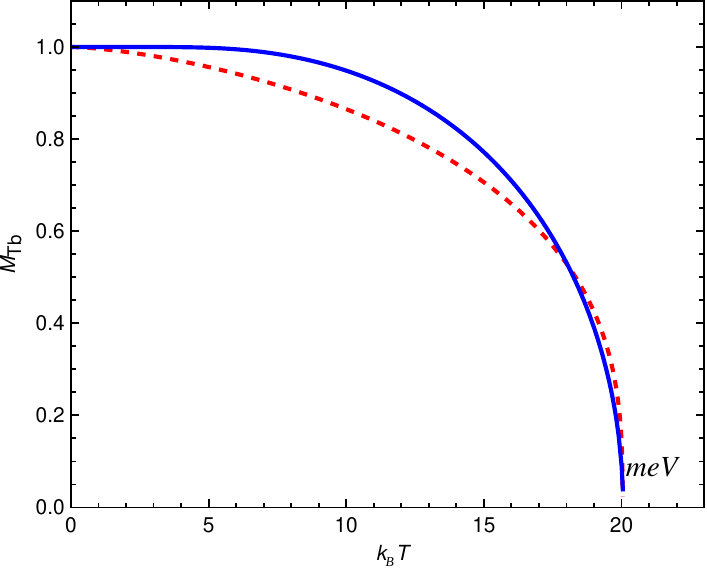}
		\caption{EFT calculated normalized magnetizations (continuous blue curve) for $hcp- Tb$ across varying temperatures compared with experimental measurements (dashed red curve).}
		\label{efttb}
	\end{figure}
	
	The exchange constant  $J_{\text{Co-Co}}$=15.9 meV \cite{sublatticemag}, $J_{\text{Tb-Tb}}$= 1.293 meV    and the EFT calculated normalized magnetizations for $hcp$-$\mathrm{Co}$ across varying temperatures compared with experimental data \cite{sublatticemag,kuz2005shape} are depicted in Fig. \ref{eftco}. Given the close correspondence between both curves, it can be concluded that the model demonstrates good agreement with $S_{Co}=1$ and
	$J_{\text{Co-Co}}$=15.9 meV  (blue continuous curve). Similarly, EFT calculated normalized magnetizations with  $S=3$ and exchange $J_{\text{Tb-Tb}}$= 1.293 meV  \cite{moujaes2015,kerr1988}  for $hcp- Tb$ across varying temperatures compared with experimental data are presented in Fig. \ref{efttb}. The exchange coefficient between $Co$ and $Tb$ i.e $J_{\text{Co-Tb}}$= -1.5 meV is taken from Shigeo Honda et al. \cite{kerr1988} . These values are then used for numerical computation to determine spin variables. By utilizing the calculated these variables along with the g factors ($g$),  which are $2.2$ and $1.5$ \cite{finley2016} for cobalt and terbium,respectively, the magnetic moment per site can be determined as a function of temperature. Moderate variations in exchange constants primarily shift the Curie and compensation temperatures while preserving the qualitative behaviour of the layer-resolved magnetization. Hence, therefore the exchange constants calculated from EFT model in accordance with experimental results has been used in the numerical computation of theoretical model.	
	
	\section{Ferrimagnetic lemellar alloy layered nanostructure between semi-infinite cobalt leads:}
	For our calculation of sublattice magnetization, we use the exchange constant values obtained from the EFT computation technique and input them into MFT-based expressions. We assume that the Co-Tb multilayer alloy nano-junction can be modeled as an hcp crystal structure with randomly oriented Co and Tb atoms. Although the actual structure might be amorphous and more complex, we consider this approximation. In their pure states, both cobalt and terbium form hcp crystal structures.
	For our study, we assume that the orientations of the atoms are uniformly random in each basal plane of the alloy nano-junction and in addition to this, we have neglected differences in atomic sizes. Each (0001) basal plane site in the hcp structure has six nearest neighbors within the same plane and six from the neighboring planes. Alternating hcp (0001) atomic planes constitute the system, with abrupt and crystalline interfaces between the leads and the layered nanostructure. By the aid of ultramodern experimental methods for Co-Tb multilayer systems, one can reduce interface roughness and regulate atomic interdiffusion and obtain a stable eutectic composition $c=0.1-0.5$.
	In order to calculate the sublattice magnetization for the individual basal atomic hcp planes of the alloy-layered [$l'll'$] nanostructures with the use of MFT, we begin with Brillouin's functions to find the different spin variables and proceed with the thermodynamic canonical spin variable which can be given as,
	
	\begin{equation}
		\begin{aligned}
			\sigma_{\alpha}^{(n')}&=\frac{2S_{\alpha} + 1}{2}\coth{\left( \frac{2S_{\alpha} + 1}{2S_{\alpha}} \frac{H_{\alpha}^{(n')}}{ k_{\mathrm{B}}T} \right)}\\&-\frac{1}{2}\coth{\left( \frac{1}{2S_{\alpha}}\frac{H_{\alpha}^{(n')}}{ k_{\mathrm{B}}T} \right)} .
			\label{brillouin}
		\end{aligned}
	\end{equation}
	For the  ${n'}$th alloy atomic plane, $\sigma_{\alpha}^{(n')}$ is thermodynamic spin variable such that $\sigma_{\alpha}^{(n')}$=$S{_{\alpha}}$ and $B{_{\alpha}}\equiv B{_{\alpha}} (S{_{\alpha}},T,H{_{\alpha}^{(n')}})$.
	Where ${\alpha}$ refers to the atomic elements, either for Co or Tb. $S{_{\alpha}}$ represents the spin of the corresponding atomic element, and $H{_{\alpha}^{(n')}}$ denotes molecular field energy for $\alpha$ element in $n'$ plane resulting from interactions with the nearest neighboring atoms (z=12) in the ${n'}$th atomic layer. $ k_{\mathrm{B}}T$ represents thermal energy given in meV.
	Taking $g{_{\alpha}^{(n')}}$ as the g factor for the element of alloy situated in ${n'}$th plane, the effective magnetic moment, $\bar{M}^{(n')}$, per site in the ${n'}$th plane is expressed as the following (in units of Bohr magnetons),
	\begin{equation}
		\bar{M}^{(n')}/ \mu_{B} = (1-c)g_{Co}^{(n')} \sigma_{Co}^{(n')} + cg_{Tb}^{(n')} \sigma_{Tb}^{(n')} .
		\label{average of M}
	\end{equation}
	\subsection{\texorpdfstring{$[\text{Co}_{1-c}\text{Tb}_{c}]_{2}[\text{Co}]_{2}[\text{Co}_{1-c}\text{Tb}_{c}]_{2}$}{[Co1-cTbc]2[Co]2[Co1-cTbc]2} alloy layered ferrimagnetic nanostructure between cobalt leads}
	This section deals with the $[2'22']$ alloy layered nano structure implanted in between the cobalt leads with $c$ and $(1-c)$ as the concentration of Tb and Co atoms. In order to obtain the molecular field energy for the Co atom within the hcp first basal plane in the nano-layered structure on the surface adjacent to cobalt lead under the MFT treatment, it is expressed as,
	\begin{align}
		&\begin{aligned}
			H_{\text{Co}}^{(1)} &= (3 \sigma_{\text{Co}}^{(B)}J_{\text{cc}}) + 6[(1-c)\sigma_{\text{Co}}^{(1)}J_{\text{cc}} + c\sigma_{\text{Tb}}^{(1)}J_{\text{ct}}] 
			\\&+3[(1-c)\sigma_{\text{Co}}^{(2)}J_{\text{cc}}+c\sigma_{\text{Tb}}^{(2)}J_{\text{ct}}] .
			\label{coH}
		\end{aligned} 
	\end{align}
	For simplification the notations for exchange interaction are denoted as $J_{\text{Co-Co}} \equiv J_{\text{cc}}$, $J_{\text{Co-Tb}}\equiv J_{\text{ct}}$ and $J_{\text{Tb-Tb}} \equiv J_{\text{tt}}$. Equivalently, we can express the molecular field energy of a Tb atom on the first hcp basal plane in the layered nanostructure at the interface with cobalt lead according to MFT, as below,
	\begin{align}
		&\begin{aligned}
			H_{\text{Tb}}^{(1)} &= (3 \sigma_{\text{Co}}^{(B)}J_{\text{ct}}) + 6[(1-c)\sigma_{\text{Co}}^{(1)}J_{\text{ct}} + c\sigma_{\text{Tb}}^{(1)}J_{\text{tt}}] 
			\\&+3[(1-c)\sigma_{\text{Co}}^{(2)}J_{\text{ct}}+c\sigma_{\text{Tb}}^{(2)}J_{\text{tt}}] .
			\label{TbH}
		\end{aligned} 
	\end{align}
	In our current conception, initial spin value for lead of pure Co atom at the interface with alloy-layered nanostructure is designated by $\sigma_{\text{Co}}^{(B)}$.
	Molecular field energy of cobalt for second basal plane of layered nanostructure at the interface with semi-infinite lead of cobalt is as follows,
	\begin{align}
		&\begin{aligned}
			H_{\text{Co}}^{(2)} &= 3[(1-c)\sigma_{\text{Co}}^{(1)}J_{\text{cc}}+c\sigma_{\text{Tb}}^{(1)}J_{\text{ct}}]  \\& +6[(1-c)\sigma_{\text{Co}}^{(2)}J_{\text{cc}} +c\sigma_{\text{Tb}}^{(2)}J_{\text{ct}}]+3\sigma_{\text{Co}}^{(3)}J_{\text{cc}}] .
			\label{coH2}
		\end{aligned} 
	\end{align}
	Likewise,we can express molecular field energy of cobalt on the second basal plane of layered nanostructure at the interface with lead of pure cobalt is as follows,
	\begin{align}
		&\begin{aligned}
			H_{\text{Tb}}^{(2)} &= 3[(1-c)\sigma_{\text{Co}}^{(1)}J_{\text{ct}}+c\sigma_{\text{Tb}}^{(1)}J_{\text{tt}}]  \\& +6[(1-c)\sigma_{\text{Co}}^{(2)}J_{\text{ct}} +c\sigma_{\text{Tb}}^{(2)}J_{\text{tt}}]+3\sigma_{\text{Co}}^{(3)}J_{\text{ct}}] .
			\label{TbH2}
		\end{aligned} 
	\end{align}
	Basal planes of hcp for $[\text {Co}]_{2}$ which is layered in between the $[\text {Co} _{1-c} \text {Tb}_{c}]_{2}$ are called as third and fourth atomic planes,respectively. With the help of symmetry attributes for $[2'22']$ layered alloy nano-junction we observed that $\sigma_{\text{Co}}^{(3)}\equiv \sigma_{\text{Co}}^{(4)}$.
	The molecular field energy for cobalt on third atomic plane is as follows,
	\begin{align}
		&\begin{aligned}
			H_{\text{Co}}^{(3)} &= 3[(1-c)\sigma_{\text{Co}}^{(2)}J_{\text{cc}}+c\sigma_{\text{Tb}}^{(2)}J_{\text{ct}}]  \ +9\sigma_{\text{Co}}^{(3)}J_{\text{cc}} .
			\label{CoH3}
		\end{aligned} 
	\end{align}
	The equations discussed above can have the following matrix form,
	\renewcommand{\arraystretch}{2} 
	\begin{equation}
		\begin{bmatrix}
			H_{\text{Co}}^{(1)} \\ 
			H_{\text{Tb}}^{(1)} \\ 
			H_{\text{Co}}^{(2)} \\ 
			H_{\text{Tb}}^{(2)} \\ 
			H_{\text{Co}}^{(3)}
		\end{bmatrix}
		=
		\begin{bmatrix}
			X_1 \\ 
			X_2 \\ 
			0 \\ 
			0 \\ 
			0
		\end{bmatrix}
		+
		\begin{bmatrix}
			s_1 & s_2 & s_3 & s_4 & s_5 \\ 
			t_1 & t_2 & t_3 & t_4 & t_5 \\ 
			u_1 & u_2 & u_3 & u_4 & u_5 \\ 
			v_1 & v_2 & v_3 & v_4 & v_5 \\ 
			w_1 & w_2 & w_3 & w_4 & w_5
		\end{bmatrix}
		\begin{bmatrix}
			\sigma_{\text{Co}}^{(1)} \\ 
			\sigma_{\text{Tb}}^{(1)} \\ 
			\sigma_{\text{Co}}^{(2)} \\ 
			\sigma_{\text{Tb}}^{(2)} \\ 
			\sigma_{\text{Co}}^{(3)} 
		\end{bmatrix}
		\label{eq:matrix_equation222} .
	\end{equation}
	From above equation we can write the coefficient matrix as,
	\begin{equation*}
		\begin{bmatrix}
			s_1 & s_2 & s_3 & s_4 & s_5 \\ 
			t_1 & t_2 & t_3 & t_4 & t_5 \\ 
			u_1 & u_2 & u_3 & u_4 & u_5 \\ 
			v_1 & v_2 & v_3 & v_4 & v_5 \\ 
			w_1 & w_2 & w_3 & w_4 & w_5
		\end{bmatrix} 
	\end{equation*}
	\begin{equation}
		\equiv \begin{bmatrix}
			2\alpha J_\text{cc} & 6c J_\text{ct} & \alpha J_\text{cc} & 3c J_\text{ct} & 0 \\
			2\alpha J_\text{ct} & 6c J_\text{tt} & \alpha J_\text{ct} & 3cJ_\text{tt} & 0 \\
			\alpha J_\text{cc} & 3c J_\text{ct} & 2\alpha J_\text{cc} & 6c J_\text{ct} & 3 J_\text{cc} \\
			\alpha J_\text{ct} & 3c J_\text{tt} & 2\alpha J_\text{ct} & 6c J_\text{tt} & 3 J_\text{ct} \\
			0 & 0 & \alpha J_\text{cc} & 3c J_\text{ct} & 9 J_\text{cc}\\
		\end{bmatrix}
		\label{coeffmatrix222} .
	\end{equation}
	From above $\alpha$ is defined as $3(1-c)$, $X_{1}$ and $X_{2}$ are $3\sigma_{\text{Co}}^{(B)} J_\text {cc}$ and $3\sigma_{\text{Co}}^{(B)} J_\text {ct}$ ,respectively.
	\begin{equation}
		\begin{bmatrix}
			\sigma_{\text{Co}}^{(1)} \\
			\sigma_{\text{Tb}}^{(1)} \\
			\sigma_{\text{Co}}^{(2)} \\
			\sigma_{\text{Tb}}^{(2)} \\
			\sigma_{\text{Co}}^{(3)}
		\end{bmatrix}
		=
		\begin{bmatrix}
			B_{\text{Co}}(S_{\text{Co}}, T, H_{\text{Co}}^{(1)}) \\
			B_{\text{Tb}}(S_{\text{Tb}}, T, H_{\text{Tb}}^{(1)}) \\
			B_{\text{Co}}(S_{\text{Co}}, T, H_{\text{Co}}^{(2)}) \\
			B_{\text{Tb}}(S_{\text{Tb}}, T, H_{\text{Tb}}^{(2)}) \\
			B_{\text{Co}}(S_{\text{Co}}, T, H_{\text{Co}}^{(3)})
		\end{bmatrix}
		\label{sigmamatrix222} .
	\end{equation}
	By using equations (\ref{brillouin})-(\ref{CoH3}), equation (\ref{eq:matrix_equation222}) represents matrix of non-linear equations which were solved for spin variables as given in equation (\ref{sigmamatrix222}).
	
	By the use of symmetry following can be simplified,
	
	\[
	\begin{bmatrix}
		\sigma_{\text{Co}}^{(1)} \\
		\sigma_{\text{Tb}}^{(1)}
	\end{bmatrix}
	\equiv
	\begin{bmatrix}
		\sigma_{\text{Co}}^{(6)} \\
		\sigma_{\text{Tb}}^{(6)}
	\end{bmatrix}, \quad
	\begin{bmatrix}
		\sigma_{\text{Co}}^{(2)} \\
		\sigma_{\text{Tb}}^{(2)}
	\end{bmatrix}
	\equiv
	\begin{bmatrix}
		\sigma_{\text{Co}}^{(5)} \\
		\sigma_{\text{Tb}}^{(5)}
	\end{bmatrix}
	\]
	\begin{equation}
		\quad
		\begin{bmatrix}
			\sigma_{\text{Co}}^{(3)}
		\end{bmatrix}
		\equiv
		\begin{bmatrix}
			\sigma_{\text{Co}}^{(4)}
		\end{bmatrix}.
	\end{equation}
	\label{symm222}
	
\begin{figure}
		\centering
		\includegraphics[width=0.8\linewidth]{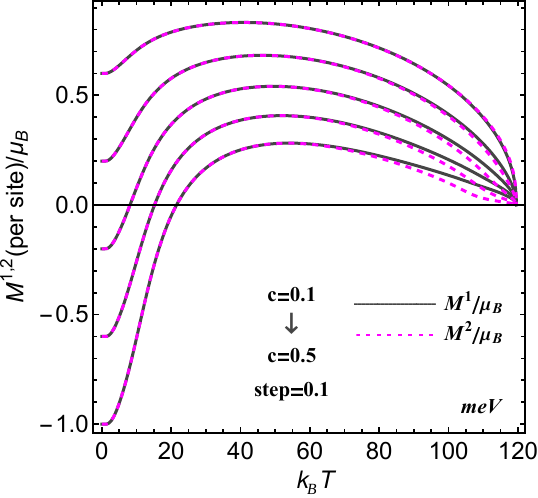}
		\caption{The magnetic moments per site, computed for atoms in the first (solid curves) and second (dashed curves) hcp basal (0001) planes of the alloy-layered ferrimagnetic nanostructure
			$\left[\text{Co}_{1-c} \text{Tb}_c\right]_2 \left[\text{Co}_2\right] \left[\text{Co}_{1-c} \text{Tb}_c\right]_2$, placed between cobalt leads,
			show only minor differences near the end of the ordered phase as temperature ($ k_{\mathrm{B}}T$) rises. The down arrows indicate how
			the site-resolved magnetic moments change with stepwise increase in $c$.} 
		\label{M12}
	\end{figure}

	Values of $\sigma_{\text{Co}}^{(1)},\sigma_{\text{Tb}}^{(1)},\sigma_{\text{Co}}^{(2)},\sigma_{\text{Tb}}^{(2)}$ and $\sigma_{\text{Co}}^{(3)}$ can be solved numerically with the help of above equations for required concentration.
	Fig. \ref{M12} has description of first and second basal hcp planes for magnetic moments as a function of temperature using equation \ref{average of M} and values of g factor for Co and Tb, which are 2.2 and 1.5, respectively, for all values of $n'$.In the figure, compensation temperature ($ k_{\mathrm{B}}T<$21 meV) can be noticed in the ferrimagnetic hcp plane for the given eutectic concentration with a range from 0.3 to 0.5. In this study compensation occurs due to the different temperature dependences of the Co and Tb sublattice magnetizations, while the antiferromagnetic Co–Tb exchange interaction keeps the two sublattices oppositely aligned. We also discuss how increasing the Tb concentration strengthens the rare-earth contribution, which leads to a shift in the compensation temperature. In addition, the interface layers have exchange environments that are different from those of the interior layers, resulting in layer-dependent magnetization.
	
	\subsection{\texorpdfstring{$[\text{Co}_{1-c}\text{Tb}_{c}]_{3}[\text{Co}]_{3}[\text{Co}_{1-c}\text{Tb}_{c}]_{3}$ alloy layered ferrimagnetic nanostructure between cobalt leads}{[Co1-cTbc]3[Co]3[Co1-cTbc]3 alloy layered ferrimagnetic nanostructure between cobalt leads}}
	By the symmetry properties,
	\begin{equation*}
		\begin{bmatrix}
			\sigma_{\text{Co}}^{(1)} \\
			\sigma_{\text{Tb}}^{(1)}
		\end{bmatrix}
		\equiv
		\begin{bmatrix}
			\sigma_{\text{Co}}^{(9)} \\
			\sigma_{\text{Tb}}^{(9)}
		\end{bmatrix}, \quad
		\begin{bmatrix}
			\sigma_{\text{Co}}^{(2)} \\
			\sigma_{\text{Tb}}^{(2)}
		\end{bmatrix}
		\equiv
		\begin{bmatrix}
			\sigma_{\text{Co}}^{(8)} \\
			\sigma_{\text{Tb}}^{(8)}
		\end{bmatrix}, \quad
	\end{equation*}
	\begin{equation}
		\begin{bmatrix}
			\sigma_{\text{Co}}^{(3)} \\
			\sigma_{\text{Tb}}^{(3)}
		\end{bmatrix}
		\equiv
		\begin{bmatrix}
			\sigma_{\text{Co}}^{(7)} \\
			\sigma_{\text{Tb}}^{(7)}
		\end{bmatrix} 
		\quad \text{and} \quad
		\begin{bmatrix}
			\sigma_{\text{Co}}^{(4)}
		\end{bmatrix}
		\equiv
		\begin{bmatrix}
			\sigma_{\text{Co}}^{(6)}
		\end{bmatrix}
		\label{symm333}.
	\end{equation}
	
\begin{figure} 
		\centering
		\includegraphics[width=0.8\linewidth]{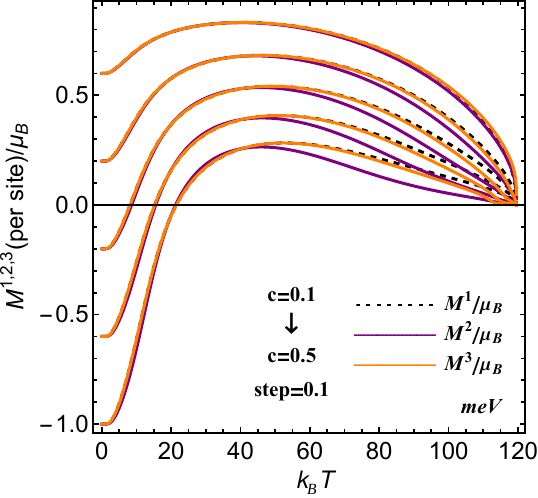}
		\caption{The computed magnetic moments per site for the first, second and third hcp basal planes of [Co${_{1-c}}$Tb${_{c}}]{_{3}}$[Co]${_{3}}$[Co${_{1-c}}$Tb${_{c}}]{_{3}}$ alloy.}
		\label{M123}
	\end{figure}

	For [3'33'] layered alloy nanostructures, there is symmetry about the origin, which is for $n=0$ in hcp plane.Equivalently, we can also represent the molecular field energy for [3'33'] like it was represented for [2'22'].
	\renewcommand{\arraystretch}{2}  
	
	\begin{equation}
		\resizebox{.8\hsize}{!}{$
			\begin{bmatrix}
				H_{\text{Co}}^{(1)} \\
				H_{\text{Tb}}^{(1)} \\
				H_{\text{Co}}^{(2)} \\
				H_{\text{Tb}}^{(2)} \\
				H_{\text{Co}}^{(3)} \\
				H_{\text{Tb}}^{(3)} \\
				H_{\text{Co}}^{(4)} \\
				H_{\text{Co}}^{(5)}
			\end{bmatrix}
			=
			\begin{bmatrix}
				X_1 \\
				X_2 \\
				0 \\
				0 \\
				0 \\
				0 \\
				0 \\
				0
			\end{bmatrix}
			+
			\begin{bmatrix}
				2\alpha J_{\text{cc}} & 6cJ_{\text{ct}} & \alpha J_{\text{cc}} & 3cJ_{\text{ct}} & 0 & 0 & 0 & 0 \\
				2\alpha J_{\text{ct}} & 6cJ_{\text{tt}} & \alpha J_{\text{ct}} & 3cJ_{\text{tt}} &  0 & 0 & 0 & 0 \\
				\alpha J_{\text{cc}} & 3cJ_{\text{ct}} & 2\alpha J_{\text{cc}} & 6cJ_{\text{ct}} & \alpha J_{\text{cc}} & 3cJ_{\text{ct}} & 0 & 0 \\
				\alpha J_{\text{ct}} & 3cJ_{\text{tt}} & 2\alpha J_{\text{ct}} & 6cJ_{\text{tt}} & \alpha J_{\text{ct}} & 3cJ_{\text{tt}} & 0 & 0 \\
				0 & 0 & \alpha J_{\text{cc}} & 3cJ_{\text{ct}} & 2\alpha J_{\text{cc}} & 6cJ_{\text{ct}} & 3J_{\text{cc}} & 0 \\
				0 & 0 & \alpha J_{\text{ct}} & 3cJ_{\text{tt}} & 2\alpha J_{\text{cc}} & 6c J_{\text{tt}} & 3J_{\text{ct}} & 0 \\
				0 & 0 & 0 & 0 & \alpha J_{\text{cc}} & 3cJ_{\text{ct}} & 6 J_{\text{cc}} & 3 J_{\text{cc}} \\
				0 & 0 & 0 & 0 & 0 & 0 & 6J_{\text{cc}} & 6J_{\text{cc}} \\
			\end{bmatrix}
			\begin{bmatrix}
				\sigma_{\text{Co}}^{(1)} \\
				\sigma_{\text{Tb}}^{(1)} \\
				\sigma_{\text{Co}}^{(2)} \\
				\sigma_{\text{Tb}}^{(2)} \\
				\sigma_{\text{Co}}^{(3)} \\
				\sigma_{\text{Tb}}^{(3)} \\
				\sigma_{\text{Co}}^{(4)} \\
				\sigma_{\text{Co}}^{(5)}
			\end{bmatrix}
			$}
		\label{eq:matrix_equation333}.
	\end{equation}
	From Eqs. \ref{coH}-\ref{CoH3} we can observe that there is a change in molecular field energies concerning temperature variations for both cobalt and terbium. Due to this variation in the molecular field energies in the whole temperature range, values of magnetizations ($M$) can be found. From the observation of Fig. \ref{M12} and Fig. \ref{M123}, we have observed that magnetization increases at low temperature and the maximum value of $M$ is reached before the loss of magnetic phase.
	As we said for [2'22'], in a similar way, we can solve non-linear equations (Eq. \ref{sigmamatrix222}) analytically to acquire magnetic moments and spin variables separately in each hcp basal plane of the multilayered alloy nanostructure.
	Where irriducible spin variables can be expressed as follows,
	\renewcommand{\arraystretch}{2}
	\begin{equation}
		\begin{bmatrix}
			\sigma_{\text{Co}}^{(1)} \\
			\sigma_{\text{Tb}}^{(1)} \\
			\sigma_{\text{Co}}^{(2)} \\
			\sigma_{\text{Tb}}^{(2)} \\
			\sigma_{\text{Co}}^{(3)}\\
			\sigma_{\text{Tb}}^{(3)}\\
			\sigma_{\text{Co}}^{(4)}\\
			\sigma_{\text{Co}}^{(5)}\\
		\end{bmatrix}
		=
		\begin{bmatrix}
			B_{\text{Co}}(S_{\text{Co}}, T, H_{\text{Co}}^{(1)}) \\
			B_{\text{Tb}}(S_{\text{Tb}}, T, H_{\text{Tb}}^{(1)}) \\
			B_{\text{Co}}(S_{\text{Co}}, T, H_{\text{Co}}^{(2)}) \\
			B_{\text{Tb}}(S_{\text{Tb}}, T, H_{\text{Tb}}^{(2)}) \\
			B_{\text{Co}}(S_{\text{Co}}, T, H_{\text{Co}}^{(3)}) \\
			B_{\text{Co}}(S_{\text{Tb}}, T, H_{\text{Tb}}^{(3)}) \\
			B_{\text{Co}}(S_{\text{Co}}, T, H_{\text{Co}}^{(4)}) \\
			B_{\text{Co}}(S_{\text{Co}}, T, H_{\text{Co}}^{(5)}) \\ 
		\end{bmatrix}
		\label{sigmamatrix333} .
	\end{equation}
	
	It was well defined earlier that for cobalt and terbium are $S{_\text{Co}}$=1 and $S{_\text{Tb}}$=3, respectively. For this system of alloy, which is ferrimagnetic, we have assumed that spin up is for cobalt and spin down is for terbium. Exchange constant for terbium is weaker than that of cobalt, even though $S_\text{Tb}$ is three times that of $S_\text{Co}$.
	
	The model for computation that we discussed can be used in general for more atomic planes, too. In general, for layered alloy system with $x,y,z \geq 1 $ [Co${_{(1-c)}}$ Tb$_{c}$]${_{x}}$ [Co]$_{y}$ [Co${_{(1-c)}}$ Tb$_{c}$]${_{z}}$ placed between semi-infinite cobalt leads.Although generalization of this computation is possible but with the escalation in the values of $x,y,z$, the consequence of this is that we get the limiting solutions for sublattice magnetizations for core atomic planes.
	
	\subsection{\texorpdfstring{$[\text{Co}_{1-c}\text{Tb}_{c}]_{2}[\text{Co}]_{3}[\text{Co}_{1-c}\text{Tb}_{c}]_{2}$ and $[\text{Co}_{1-c}\text{Tb}_{c}]_{3}[\text{Co}]_{2}[\text{Co}_{1-c}\text{Tb}_{c}]_{3}$ alloy layered ferrimagnetic nanostructure between cobalt leads}{[Co1-cTbc]2[Co]3[Co1-cTbc]2 and [Co1-cTbc]3[Co]2[Co1-cTbc]3 alloy layered ferrimagnetic nanostructure between cobalt leads}}
	The $2'32'$ and $3'23'$  Co-Tb layered structures show little different magnetic behaviors due to their unique stacking patterns. The molecular field energy can be calculated in the equivalent way as $2'22'$ and $3'33'$. 
	The following are matrix form representations for molecular field energy of the $2'32'$ ( Eq. \ref{eq:matrix_equation232}) and $3'23'$ (Eq. \ref{eq:matrix_equation323}),
	
	\begin{equation}
		\resizebox{.8\hsize}{!}{$
			\begin{bmatrix}
				H_{\text{Co}}^{(1)} \\ H_{\text{Tb}}^{(1)} \\ H_{\text{Co}}^{(2)} \\ 
				H_{\text{Tb}}^{(2)} \\ H_{\text{Co}}^{(3)} \\ H_{\text{Co}}^{(4)}
			\end{bmatrix}
			=
			\begin{bmatrix}
				X_1 \\ X_2 \\ 0 \\ 0 \\ 0 \\ 0
			\end{bmatrix}
			+
			\begin{bmatrix}
				2\alpha J_\text{cc} & 6c J_\text{ct} & \alpha J_\text{cc} & 3c J_\text{ct} & 0 &0\\
				2\alpha J_\text{ct} & 6c J_\text{tt} & \alpha J_\text{ct} & 3cJ_\text{tt} & 0&0 \\
				\alpha J_\text{cc} & 3c J_\text{ct} & 2\alpha J_\text{cc} & 6c J_\text{ct} & 3 J_\text{cc} &0\\
				\alpha J_\text{ct} & 3c J_\text{tt} & 2\alpha J_\text{ct} & 6c J_\text{tt} & 3 J_\text{ct} &0\\
				0 & 0 & \alpha J_\text{cc} & 3c J_\text{ct} & 6 J_\text{cc}& 3J_\text{cc}\\
				0& 0&0&0&6 J_\text{cc}& 6 J_\text{cc}\\
				
			\end{bmatrix}
			\begin{bmatrix}
				\sigma_{\text{Co}}^{(1)} \\ \sigma_{\text{Tb}}^{(1)} \\ \sigma_{\text{Co}}^{(2)} \\ 
				\sigma_{\text{Tb}}^{(2)} \\ \sigma_{\text{Co}}^{(3)} \\ \sigma_{\text{Co}}^{(4)}
			\end{bmatrix}
			$}
		\label{eq:matrix_equation232} ,
	\end{equation}
	
	\begin{equation}
		\begin{bmatrix}
			\sigma_{\text{Co}}^{(1)} \\
			\sigma_{\text{Tb}}^{(1)} \\
			\sigma_{\text{Co}}^{(2)} \\
			\sigma_{\text{Tb}}^{(2)} \\
			\sigma_{\text{Co}}^{(3)}\\
			\sigma_{\text{Co}}^{(4)}\\
		\end{bmatrix}
		=
		\begin{bmatrix}
			B_{\text{Co}}(S_{\text{Co}}, T, H_{\text{Co}}^{(1)}) \\
			B_{\text{Tb}}(S_{\text{Tb}}, T, H_{\text{Tb}}^{(1)}) \\
			B_{\text{Co}}(S_{\text{Co}}, T, H_{\text{Co}}^{(2)}) \\
			B_{\text{Tb}}(S_{\text{Tb}}, T, H_{\text{Tb}}^{(2)}) \\
			B_{\text{Co}}(S_{\text{Co}}, T, H_{\text{Co}}^{(3)}) \\ B_{\text{Co}}(S_{\text{Co}}, T, H_{\text{Co}}^{(4)}) \\
		\end{bmatrix}
		\label{sigmamatrix232} .
	\end{equation}
	
	\begin{equation}
		\resizebox{.8\hsize}{!}{$
			\begin{bmatrix}
				H_{\text{Co}}^{(1)} \\ H_{\text{Tb}}^{(1)} \\ H_{\text{Co}}^{(2)} \\ H_{\text{Tb}}^{(2)} \\
				H_{\text{Co}}^{(3)} \\ H_{\text{Tb}}^{(3)} \\ H_{\text{Co}}^{(4)}
			\end{bmatrix}
			=
			\begin{bmatrix}
				X_1 \\ X_2 \\ 0 \\ 0 \\ 0 \\ 0 \\ 0
			\end{bmatrix}
			+
			\begin{bmatrix}
				2\alpha J_{\text{cc}} & 6cJ_{\text{ct}} & \alpha J_{\text{cc}} & 3cJ_{\text{ct}} & 0 & 0 & 0 \\
				2\alpha J_{\text{ct}} & 6cJ_{\text{tt}} & \alpha J_{\text{ct}} & 3cJ_{\text{tt}} & 0 & 0 & 0 \\
				\alpha J_{\text{cc}} & 3cJ_{\text{ct}} & 2\alpha J_{\text{cc}} & 6cJ_{\text{ct}} & \alpha J_{\text{cc}} & 3cJ_{\text{ct}} & 0  \\
				\alpha J_{\text{ct}} & 3cJ_{\text{tt}} & 2\alpha J_{\text{ct}} & 6cJ_{\text{tt}} & \alpha J_{\text{ct}} & 3cJ_{\text{tt}} & 0 \\
				0 & 0 & \alpha J_{\text{cc}} & 3cJ_{\text{ct}} & 2\alpha J_{\text{cc}} & 6cJ_{\text{ct}} & 3J_{\text{cc}}  \\
				0 & 0 & \alpha J_{\text{ct}} & 3cJ_{\text{tt}} & 2\alpha J_{\text{ct}} & 6c J_{\text{tt}} & 3J_{\text{ct}}  \\
				0 & 0 & 0 & 0 & \alpha J_{\text{cc}} & 3cJ_{\text{tt}} & 9J_{\text{cc}}
			\end{bmatrix}
			\begin{bmatrix}
				\sigma_{\text{Co}}^{(1)} \\ \sigma_{\text{Tb}}^{(1)} \\ \sigma_{\text{Co}}^{(2)} \\ 
				\sigma_{\text{Tb}}^{(2)} \\ \sigma_{\text{Co}}^{(3)} \\ \sigma_{\text{Tb}}^{(3)} \\ \sigma_{\text{Co}}^{(4)}
			\end{bmatrix}
			$}
		\label{eq:matrix_equation323} ,
	\end{equation}
	
	\begin{equation}
		\begin{bmatrix}
			\sigma_{\text{Co}}^{(1)} \\
			\sigma_{\text{Tb}}^{(1)} \\
			\sigma_{\text{Co}}^{(2)} \\
			\sigma_{\text{Tb}}^{(2)} \\
			\sigma_{\text{Co}}^{(3)}\\
			\sigma_{\text{Tb}}^{(3)}\\
			\sigma_{\text{Co}}^{(4)}\\
		\end{bmatrix}
		=
		\begin{bmatrix}
			B_{\text{Co}}(S_{\text{Co}}, T, H_{\text{Co}}^{(1)}) \\
			B_{\text{Tb}}(S_{\text{Tb}}, T, H_{\text{Tb}}^{(1)}) \\
			B_{\text{Co}}(S_{\text{Co}}, T, H_{\text{Co}}^{(2)}) \\
			B_{\text{Tb}}(S_{\text{Tb}}, T, H_{\text{Tb}}^{(2)}) \\
			B_{\text{Co}}(S_{\text{Co}}, T, H_{\text{Co}}^{(3)}) \\
			B_{\text{Co}}(S_{\text{Tb}}, T, H_{\text{Tb}}^{(3)}) \\
			B_{\text{Co}}(S_{\text{Co}}, T, H_{\text{Co}}^{(4)}) \\
		\end{bmatrix}
		\label{sigmamatrix323} .
	\end{equation}
	
	\begin{figure} [b]
		\centering
		\includegraphics[width=0.8\linewidth]{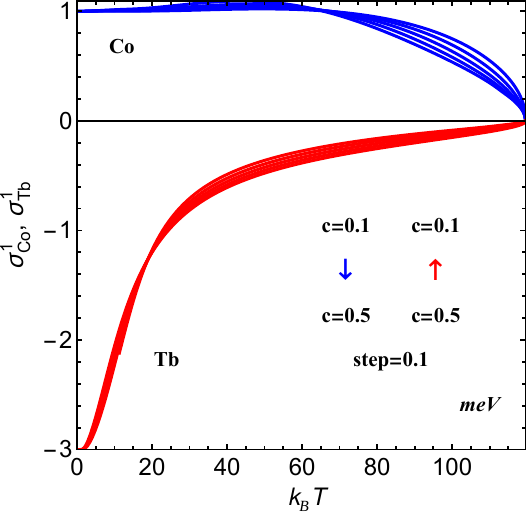}
		\caption{The graph shows the change in the spin variables for $2'32'$ alloy layered naojunction as a function of temperature.}
		\label{sigma232}
	\end{figure}
	\begin{figure}
		\centering
		\includegraphics[width=0.8\linewidth]{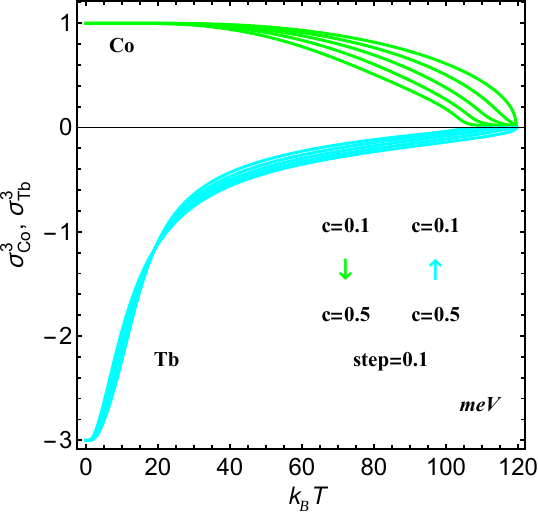}
		\caption{The graph shows the change in spin variables of $3'23'$ as a function of temperature. }
		\label{sigma323}
	\end{figure}
	Figs. \ref{sigma232} and \ref{sigma323} illustrates computed spin variables for 2'32' and 3'23' layered structure,respectively.These trends are compatible with theoretical predictions under the effective field theory approach, in which thicker Co-layers suppress the rare-earth effect and stabilize the overall magnetization profile.
	\subsection{\texorpdfstring{$[\text{Co}_{0.6}\text{Tb}_{0.4}]_{l'}[\text{Co}]_{l}[\text{Co}_{0.6}\text{Tb}_{0.4}]_{l'}$ alloy layered nanostructure between cobalt leads}{[Co0.6Tb0.4]l' [Co]l [Co0.6Tb0.4]l' alloy layered nanostructure between cobalt leads}}
	The concept for limiting solutions of spin variables and effective magnetic moments per site is illustrated in this section. 
	\begin{figure}
		\centering
		\includegraphics[scale=0.6]{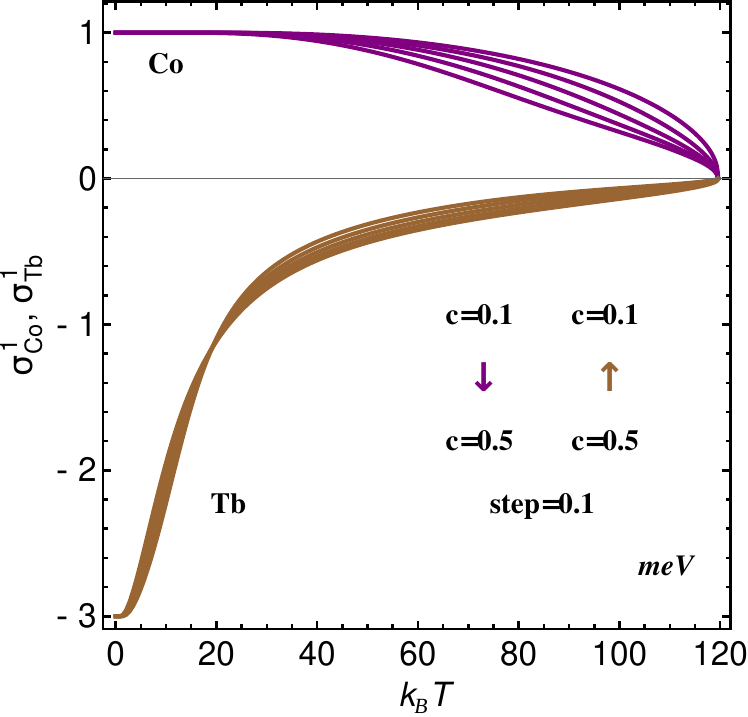} \\
		8(a)\\
		\vspace{2mm}
		\includegraphics[scale=0.60]{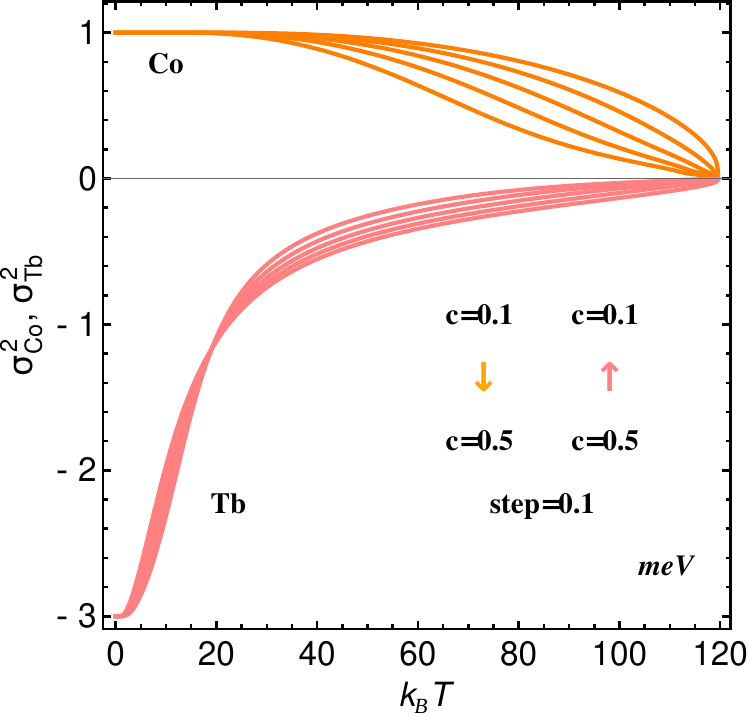}\\
		8(b)
		\caption{Computed spin variables for Co and Tb sites on the 1st (a) and 2nd (b) (0001) hcp basal planes of alloy-layered nanostructure }
		\label{sigma 222.333}
	\end{figure}
	In this section we have provided description of  the computed values of spin variables ($\sigma_{\text{Co}}^{(n')} $ and $\sigma_{\text{Tb}}^{(n')}$) for value of $n'$ = 1 and 2 as in Fig. \ref{sigma 222.333} . The plot provides the value of these spin variables as a function of temperature for a step difference of 0.1 at $c=0.1-0.5$.
	\begin{figure}
		\centering
		\includegraphics[scale=0.60]{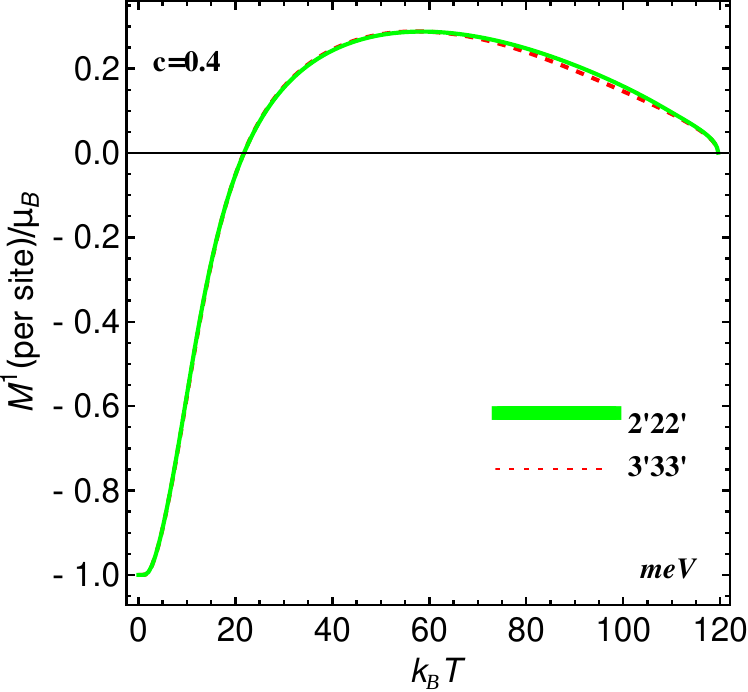}\\
		9(a) \\
		\vspace{2mm}  
		\includegraphics[scale=0.60]{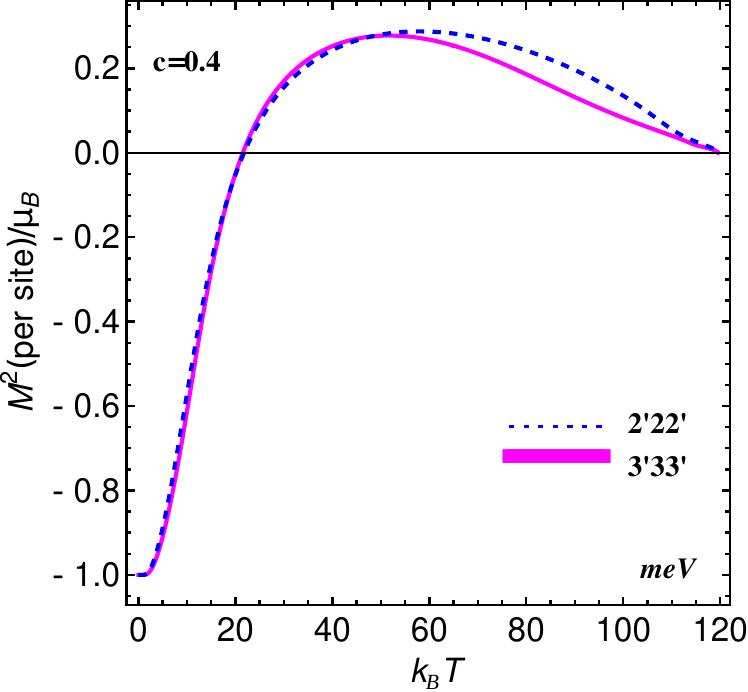}\\
		9(b)
		\caption{9(a) and 9(b) represent effective magnetic moments per site for $n'=1$ and $n'=2$, respectively, for $[\text {Co}_{0.4}]_{2'}$ [Co]$_{2}$ [Co$_{0.6}$ Tb$_{0.4}]_{2'}$ as well as $[\text{Co}_{0.4}]_{3'}$ [Co]$_{3}$ [Co$_{0.6}$ Tb$_{0.4}]_{3'}$.}
		\label{effMag}
	\end{figure}
	We analyze the alloy layered nano-junction for $c$ = 0.4.The combined EFT-MFT model is applied here to determine spin variables as well as the magnetic moments per site for the chosen concentration, which is 0.4. The determination of spin variables ($\sigma_{\text{Co}}^{(n')}$ and $\sigma_{\text{Co}}^{(n')}$) and magnetic moments is performed for layered alloy nanostructure for each hcp basal planes which is function of temperature,$c$ and thickness($l$=2 and 3). We have computed magnetic moments in the units of bohr magnetons.
	In Fig. \ref{effMag} we have discussed about magnetic moments as function of temperature for each site for $n'=1$ [Co${_{0.6}}$ Tb$_{0.4}$]${_{2'}}$ [Co]$_{2}$ [Co${_{0.6}}$ Tb$_{0.4}$]${_{2'}}$ and $n'=2$ [Co${_{0.6}}$ Tb$_{0.4}$]${_{3'}}$ [Co]$_{3}$ [Co${_{0.6}}$ Tb$_{0.4}$]${_{3'}}$ in figures (a) and  (b), 
	respectively .It was noticed that for $n'=1$ magnetic moment was not much diversified with an increase in thickness ($l'$), which is clearly evident with Eq. \ref{eq:matrix_equation333} as it does not vary much with $l'$. In contradiction to this, for $n'=2$ there is variation in the magnetic moments with variation in thickness (figure (b)). This can be noticed at $ k_{\mathrm{B}}T\approx 15$ meV up to room temperature, which is about 26 meV.
	By this we also expected $\sigma _{Co}^{(4)}$ $\leq$ $\sigma _{Co}^{(5)}$ with respect to the ferrimagnetic alloy nano-junction [Co${_{0.6}}$ Tb$_{0.4}$]${_{3'}}$ [Co]$_{3}$ [Co${_{0.6}}$ Tb$_{0.4}$]${_{3'}}$ placed between the cobalt leads. If we dicuss about ferrimagnetic alloy system [Co${_{0.6}}$ Tb$_{0.4}$]${_{2'}}$ [Co]$_{2}$ [Co${_{0.6}}$ Tb$_{0.4}$]${_{2'}}$ placed in between the cobalt leads then, $\sigma _{Co}^{(3)}$ for this system is found to be less than and equal to the  $\sigma _{Co}^{(4)}$ and $\sigma _{Co}^{(5}$, which are the spin variables with respect to 3'33' system. 
	Interestingly, the behavior of the $[2'22']$ layered structure, featuring $n' = 3$ pure cobalt planes, is found to closely resemble that of the $[3'33']$ system with $n' = 4$ pure cobalt planes. 
	This observation highlights that increasing the number of cobalt planes only slightly alters the magnetization profile for these specific configurations. 
	These findings align with theoretical expectations, showing that as the thickness of the ferrimagnetic alloy nano-junction increases, the computed magnetization values exhibit a limiting behavior. 
	
	\begin{figure}
		\centering
		\includegraphics[width=0.8\linewidth]{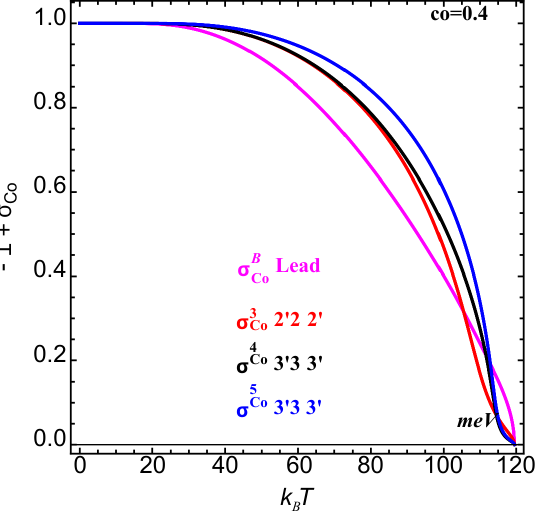}
		\caption{The calculated spin variables $\sigma_{\text{Co}}^{(n')}$ and their detailed variation with temperature.}
		\label{varatc=0.4}
	\end{figure}
	In Fig. \ref{varatc=0.4} temperature variation on the nominal cobalt hcp basal planes within the alloy-layered ferrimagnetic nanostructures $[ \text{Co}_{0.6} \text{Tb}_{0.4} ]_{2'} [\text{Co}]_{2} [ \text{Co}_{0.6} \text{Tb}_{0.4} ]_{2'}$ and $[ \text{Co}_{0.6} \text{Tb}_{0.4} ]_{3'} [\text{Co}]_{3} [ \text{Co}_{0.6} \text{Tb}_{0.4} ]_{3'}$ are presented. These are compared by varying the temperature of spin variable $\sigma_{\text{Co}}^{(B)}$ on cobalt leads.
	\begin{table*}[ht]
		\tbl{Spin variable $\sigma_{\text{Tb}} = \langle S_{\text{Tb}} \rangle$ for Tb sites on atomic planes 1, 2, \dots, $l$ in the $2'22'$ and $3'33'$ nanojunction systems between cobalt leads at eutectic composition ($c \geq 0.5$) and $T=300\ \text{K}$.}
		{
			\renewcommand{\arraystretch}{1.2}
			\setlength{\tabcolsep}{6pt}
			\begin{tabular}{c c c c c c c c c c c}
				\toprule
				\multicolumn{11}{c}{\textbf{Spin variables $\sigma_{\text{Tb}}$}} \\
				\colrule
				& \multicolumn{4}{c}{\textbf{2'22'}} & \multicolumn{6}{c}{\textbf{3'33'}} \\
				\cmidrule(lr){2-5} \cmidrule(lr){6-11}
				\textbf{Concentration $c$} & \multicolumn{2}{c}{$[\text{Co}_{1-c}\text{Tb}_c]_2$} & \multicolumn{2}{c}{$[\text{Co}_{1-c}\text{Tb}_c]_2$} &
				\multicolumn{2}{c}{$[\text{Co}_{1-c}\text{Tb}_c]_3$} & \multicolumn{2}{c}{$[\text{Co}_{1-c}\text{Tb}_c]_3$} & \multicolumn{2}{c}{$[\text{Co}_{1-c}\text{Tb}_c]_3$} \\
				& $_L$ & $_R$ & $_L$ & $_R$ & $_L$ & $_R$ & $_L$ & $_R$ & $_L$ & $_R$ \\
				\colrule
				0.1 & -0.87 & -0.87 & -0.87 & -0.87 & -0.87 & -0.87 & -0.87 & -0.87 & -0.87 & -0.87 \\
				0.2 & -0.85 & -0.85 & -0.85 & -0.85 & -0.86 & -0.84 & -0.86 & -0.86 & -0.84 & -0.86 \\
				0.3 & -0.83 & -0.83 & -0.83 & -0.83 & -0.82 & -0.81 & -0.83 & -0.82 & -0.81 & -0.83 \\
				0.4 & -0.81 & -0.81 & -0.81 & -0.81 & -0.78 & -0.78 & -0.81 & -0.78 & -0.78 & -0.81 \\
				0.5 & -0.78 & -0.78 & -0.78 & -0.78 & -0.74 & -0.74 & -0.78 & -0.74 & -0.74 & -0.78 \\
				\botrule
			\end{tabular}
		}
		\label{table spinvariables}
	\end{table*}
	This trend can be attributed to the influence of the semi-infinite cobalt leads and the structural constraints of the layered systems.  
	Overall, the results emphasize how both the temperature and the structural arrangement of the ferrimagnetic alloy significantly influence its magnetic properties. We would like to emphasize that fundamental physical variables, such as the exchange interactions and the sublattice magnetizations of both cobalt and terbium sites, play a crucial role in the calculations of spin dynamics for magnetic nanostructures. These variables are essential for studying magnon transport, particularly in the field of Magnonics, including ballistic magnon transport. In the following table, we present an example of the calculated spin variables: $\langle S_{\text{Co}} \rangle$ is constant (equal to 1) throughout the system, while $\langle S_{\text{Tb}} \rangle$ varies across the identified atomic planes 1, 2, \dots, $l$ for the layered nanostructures $[\text{....Co}] [2'22'] [\text{Co....}]$ and $[\text{.....Co}] [3'33'] [\text{Co.....}]$, embedded between cobalt leads. These calculations are carried out at room temperature (300 K) for the stable eutectic compositions, using a combined approach of EFT and  MFT.
	\section{Summary and Conclusions:}
	In this study, we apply a combined approach of EFT and MFT to investigate magnetic properties of cobalt and terbium atomic sites within a ferrimagnetic [Co${_{1-c}}$Tb${_{c}}]{_{l}}$[Co]${_{l'}}$[Co${_{1-c}}$Tb${_{c}}]{_{l'}}$ alloy nano-junction, positioned between leads of pure cobalt.Further, results were analyzed as function of temperature, eutectic composition of nano-junction ($c \leq 0.5$), and thickness $n$ across a range of concentrations that ensure the system's stability. This model is versatile and can be adapted to explore other alloy combinations and nanostructures. To be precise, we computed sublattice magnetization for cobalt and terbium sites with respect to each hcp basal plane (0001)
	using a combination of EFT and MFT methods.
	We utilize the Ising EFT method, well-known for its flexibility, to determine the exchange constants for Co and Tb crystals in their ordered magnetic phase, by cross-referencing with experimental data. The method evaluates various spin contributions, including exchange interactions, anisotropy, and Zeeman terms, using spin values for cobalt and terbium. We also computed spin variables which are provided by table 1 these values were obtained using combined EFT and MFT methods. The spin variable $\sigma_{\text{Co}} = \langle S_{\text{Co}} \rangle$ for Co sites throughout is found to be $1$ at room temperature.
	
	Moreover, we compute sub-lattice magnetizations for both Co and Tb sites, as well as magnetic moments per site in each nano-junction layer, using stable eutectic compositions. Given that the ferrimagnetic system in this study is ultra-thin and structurally different from previously studied bulk systems, drawing meaningful comparisons between the sub-lattice magnetizations obtained here and the overall magnetization calculated through the MFT method is not straightforward. This difficulty arises from the lack of adjustable parameters in our calculations.
	
	By substitution of the results obtained using MFT for alloy lamellar ferrimagnetic nanostructure and seeding them with the EFT computations for cobalt leads, starting from interface and extending inwards, the sub-lattice magnetizations for both the cobalt and terbium sites within nanostructure were calculated.
	
	The intrinsic spins of Co and Tb, along with the determined exchange constants within the nano-junction layers, are key factors in analyzing the spin dynamics and quantum transport of magnons across these ferrimagnetic junctions. This research provides new insights into the behavior of such nano-junctions, unlocking the opportunities for advanced technological applications.
    \section{Acknowledgements} One of the authors (V.A.) gratefully acknowledges the financial support received through the Core Research Grant No. CRG/2023/001573 from the Science and Engineering Research Board (SERB), Anusandhan National Research Foundation (ANRF), Government of India.


\end{document}